\documentclass[12pt]{iopart}
\usepackage[utf8]{inputenc}
\usepackage{amssymb}
\expandafter\let\csname equation*\endcsname\relax
\expandafter\let\csname endequation*\endcsname\relax
\usepackage{amsmath}
\usepackage{graphicx}
\usepackage{color}

\begin{document}

\title[$N$-body methods for relativistic cosmology]{$N$-body methods for relativistic cosmology}

\author[Julian Adamek, Ruth Durrer and Martin Kunz]{Julian Adamek${}^1$, Ruth Durrer${}^1$ and Martin Kunz${}^{1,2}$}

\address{${}^1$ D\'epartement de Physique Th\'eorique \& Center for Astroparticle Physics,\\
~\, Universit\'e de Gen\`eve, Quai E.\ Ansermet 24, 1211 Gen\`eve 4, Switzerland\\
${}^2$ African Institute for Mathematical Sciences,\\
~\, 6 Melrose Road, Muizenberg 7945, South Africa}

\ead{julian.adamek@unige.ch, ruth.durrer@unige.ch, martin.kunz@unige.ch}

\begin{abstract}
We present a framework for general relativistic $N$-body simulations in the regime of weak gravitational fields.
In this approach, Einstein's equations are expanded in terms of metric perturbations about a Friedmann-Lema\^itre background,
which are assumed to remain small. The metric perturbations themselves are only kept to linear order, but we keep
their first spatial derivatives to second order and treat their second spatial derivatives as well as sources of stress-energy
fully non-perturbatively.
The evolution of matter is modelled by an $N$-body ensemble which can consist of free-streaming nonrelativistic (e.g.\ cold dark matter) or relativistic particle species
 (e.g.\ cosmic neutrinos), but the framework is fully general and also allows for other sources of stress-energy, 
 in particular additional relativistic sources like modified-gravity models or topological defects.
We compare our method with
the traditional Newtonian approach and argue that relativistic methods are conceptually more robust and flexible,
at the cost of a moderate increase of numerical difficulty.
However, for a $\Lambda$CDM cosmology, where nonrelativistic matter is the only source of perturbations, the relativistic
corrections are expected to be small. We quantify this statement by extracting post-Newtonian estimates from
Newtonian $N$-body simulations.
\end{abstract}

\pacs{98.80.Jk, 04.25.D-, 04.25.Nx}

\submitto{\CQG}

\maketitle

\section{Introduction}
The distribution of matter in the Universe is one of the main observables in cosmology.
Exquisite measurements of the cosmic microwave background anisotropies \cite{Ade:2013zuv} give convincing evidence
that at early times the Universe started out almost homogeneous and isotropic with only
small initial fluctuations. Later, and on sufficiently small scales, these
fluctuations have grown to form the galaxies, clusters, filaments and voids we observe today. 
On very large scales fluctuations are still small and can be described reasonably well with relativistic linear perturbation theory.
On smaller scales where non-linearities are important but structures are still much larger than their Schwarzschild radius, we 
expect that the dynamics of nonrelativistic matter
is well approximated by Newtonian gravity, and we usually calculate gravitational clustering
with Newtonian $N$-body simulations.

Already to satisfy internal consistency, the Newtonian treatment requires the existence of an appropriate background solution
in the Newtonian theory. It is well known that the Newtonian analog of Friedmann's equations only allow for three types of components
which scale as cold dark matter, curvature, and a cosmological constant, respectively. Newtonian simulations are therefore intrinsically
limited to the domain of $\Lambda$ cold dark matter ($\Lambda$CDM) cosmologies. One can, of course, impose some arbitrary evolution of
the scale factor from the outset, but it is then by no means guaranteed that the results obtained this way are close to the solution of
General Relativity which one tries to emulate. A consistent treatment requires a fully relativistic framework like the one outlined here.

At any rate, as surveys become larger and measurements more precise, neither the linear perturbation calculations nor the Newtonian approximations may
be sufficiently accurate in the intermediate regime. Furthermore, the Newtonian treatment is no longer justified if there are any
relativistic sources of perturbations, and therefore its use precludes many interesting extensions or alternatives to the $\Lambda$CDM
standard cosmological model. At the dawn of upcoming large surveys like Euclid\footnote{www.euclid-ec.org} \cite{Laureijs:2011gra}, SKA\footnote{www.skatelescope.org} \cite{Carilli:2004nx}
and LSST\footnote{www.lsst.org} \cite{Abell:2009aa}, able to test our theory of gravity and possibly providing the key to understanding the nature of dark energy, we think
it is time to take our numerical techniques to the next level.

In this review we describe a framework for numerical $N$-body simulations which includes the relativistic corrections possibly
relevant for cosmology. It bridges the gap between relativistic linear perturbation theory and the Newtonian limit, but we take
into account terms which can be relevant
beyond these two limits, i.e.\ at intermediate scales or in the presence of evolving relativistic sources.
Contrary to Newtonian simulations which have only a single gravitational potential, our relativistic
approach directly evolves all six
physical degrees of freedom of the metric: the two scalar potentials, i.e.\ the Bardeen potentials $\Phi$ and $\Psi$, the two 
vector degrees of freedom and the
two helicities of gravitational waves.
The knowledge of the full metric allows to compute observables like distances, weak lensing or
the observed galaxy distribution correctly and self-consistently,
either by directly ray-tracing through the simulation volume or  by extracting the power spectra
and/or unequal-time correlators
of the metric perturbations and feeding them into the perturbative expressions which can be derived for some of those observables, see e.g.~\cite{Bonvin:2005ps,Schmidt:2012ne,DiDio:2012bu,Jeong:2013psa,Schmidt:2013gwa}. Working with unequal time correlators
obtained from simulations has been successfully applied in the past for models with topological defects and other scaling seeds~\cite{Durrer:2001cg}.

We explain the code components necessary for the simulation of
a $\Lambda$CDM cosmology but
an extension to other
models is straightforward whenever a
numerical treatment of their additional equations of motion can be provided.

The framework presented here works under following set of assumptions:
\begin{enumerate}
 \item\label{a:FLRW} Spacetime, 
on the scales of interest, can be described by a perturbed
Friedmann-Lema\^itre-Robertson-Walker (FLRW) metric. Furthermore, the spatial curvature
of the background model will be set to zero, mainly in order to allow for a simple tessellation
in terms of a cubic lattice, see \sref{sec:numerics}. This is a good approximation if
the curvature scale is much larger than the scale of the horizon, a premise that has strong observational support \cite{Ade:2013zuv}.

 \item\label{a:pert} The perturbations of the metric -- but not necessarily their gradients and curvature -- remain small on the scales of interest,
 such as to allow for a controlled perturbative expansion in these variables. Note that this does not imply that the perturbations
in the stress-energy tensor have to remain small; for instance, the gravitational potential of the sun at its surface is
$\sim 2 \times 10^{-6}$ despite the fact that its density is $\sim 10^{30}$ times larger than the mean
density of the Universe. We will therefore allow for fully non-perturbative sources of stress-energy.

 \item\label{a:smooth} We are not interested in localized strong-field regions, like the vicinity of black holes and neutron
stars, and we assume that one is allowed to smooth out the small scale matter distribution without introducing large
errors in the large scale dynamics. Even though there is no rigorous proof that this should work, there is also
absolutely no indication to the contrary. It would be very surprising if the gravitational dynamics of a galaxy that
contains some neutron stars would be completely different from the one of a galaxy that contains only ordinary stars. One, albeit incomplete,
explanation of the fact that local strong fields do not affect the field at much larger scales may be Birkhoff's theorem which also holds
approximately in non spherically symmetric and non vacuum situations~\cite{Ellis:2013dla,Goswami:2012jf}.

 \item\label{a:shortwave} Observational evidence strongly supports the proposition that scales close to or beyond the horizon are in the
 linear regime at any time. We will use this assumption to justify an expansion scheme that is well adapted to
 this situation. To be specific, we recognize that metric perturbations can contain fluctuations of short wavelengths
 while maintaining their small amplitude. Each spatial derivative effectively imposes a factor proportional to the inverse
 length scale, which means that terms containing such derivatives become more important at small scales. This leads to
 the notion of so-called ``shortwave corrections,'' see \sref{sec:shortwave}.
\end{enumerate}
Let us finally point out that we do not take a quasistatic or sub-horizon limit and we make a priori no assumptions about the nature of
the sources of stress-energy apart from the fact that they give rise to metric perturbations of small amplitude only.
However, given the central role it takes in the standard model, in this article we explicitly consider cold dark
matter as a source and only comment briefly on other possibilities.

The paper is organized as follows: In the next section we
introduce the relevant {metric and matter} degrees of freedom {as well as} the notation used throughout this review. In~\sref{sec:einstein} we write  the
equations to be solved, first at linear order in the metric perturbations
and then including the
relevant nonlinear corrections. In \sref{sec:numerics} we give more details about the numerical implementation of the problem and in
\sref{sec:prelim} we present and discuss some preliminary results. The review concludes in \sref{sec:outlook} with a list of what still has to be done
including some studies which are under way.

\section{Choice of variables}\label{sec:dof}

\subsection{The metric field}
\label{ssec:metric}

Under assumption (\ref{a:FLRW}) and adopting longitudinal gauge, the line element takes the form
\begin{equation}
 \label{e:metric}
 \rmd s^2 = a^2\!(\tau) \left[-\left(1 + 2 \Psi\right) \rmd\tau^2 - 2 B_i \rmd x^i \rmd\tau
 + \left(1 - 2 \Phi\right) \delta_{ij} \rmd x^i \rmd x^j + h_{ij} \rmd x^i \rmd x^j\right] \, ,
\end{equation}
where $a$ denotes the scale factor of the FLRW background, $x^i$ are comoving Cartesian coordinates
on the spacelike hypersurfaces, and $\tau$ is  conformal time. We shall use notation such
that Latin indices take values $1$, $2$, $3$, while Greek indices run over all four spacetime
directions, $0$, $1$, $2$, $3$. A prime denotes partial derivative with respect to $\tau$,
and we use the shorthands $f_{,i} \doteq \partial f / \partial x^i$ and
$\Delta f \doteq \delta^{ij} f_{,ij}$ for the derivatives on spacelike hypersurfaces. As usual,
the sum over repeated indices is implied. Finally, we use parentheses and square brackets to
indicate symmetrization and antisymmetrization of indices, respectively.

The above line element is supplemented by the gauge conditions
$\delta^{ij}B_{i,j} = \delta^{ik}h_{ij,k} = \delta^{ij}h_{ij} = 0$ which fixes the gauge such that
$B_i$ is a spin-1 (vector) and $h_{ij}$ is a spin-2 (tensor) perturbation. The intuitive interpretation
of these perturbation variables is that $\Phi$ and $\Psi$ are the gravitational potentials which govern
both the lensing of light rays and the clustering of matter, $B_i$ is responsible for frame-dragging
effects, and $h_{ij}$ describes gravitational waves. We will later see that the motion of particles
in the nonrelativistic regime is only sensitive to $\Psi$ and $B_i$, while relativistic species (like
photons) react to all the perturbations.

Let us comment on the residual gauge freedom in the scalar sector. There are three free choices
left which need to be fixed by convention. Firstly, the spatial coordinates can be rescaled to adjust the
value of the scale factor $a$ to any particular value at one instant in time. The usual convention is to
set $a = 1$ today. Secondly, one can always make a reparametrization of time. Observables will not change,
for instance, if one uses proper time instead of conformal time. More importantly, the homogeneous mode of
$\Psi$ and its time dependence is degenerate with the choice of time parametrization. A possible
convention is therefore to set this homogeneous mode identical to zero at all times. Thirdly, in a similar
fashion, the homogeneous mode of $\Phi$ and its time evolution is degenerate with $a$. This gives us a
certain freedom how to solve Friedmann's equations for $a$. If one wants to make sure that the
homogeneous mode of $\Phi$ remains zero, one has to consistently take into account the 
effect of perturbations on the background evolution. The presence of perturbations will slightly change
the equation of state for the background. For instance, peculiar velocities in the dark matter component
give rise to an average non-zero kinetic energy density and pressure which are not present in an unperturbed Friedmann
model (see e.g.\ \cite{Ballesteros:2011cm}). Taking these tiny effects on the evolution of $a$ into account can be numerically expensive,
because they have to be extracted  continuously from the perturbations. But we are
offered an alternative: we can choose to take a simple model for the background evolution, for
instance by adopting a fixed, approximate equation of state, and to allow for the freedom to have a non-vanishing
homogeneous mode in $\Phi$ which accounts for the inaccuracy of our background model. Einstein's equations
will then ensure that the combined evolution of $a$ and the homogeneous mode of $\Phi$ is
unique, and observables do not depend on the choice of the background model, as long as the homogeneous mode of $\Phi$ remains small.
Moreover, we can use the homogeneous mode of $\Phi$ as a diagnostic to check the accuracy of the background model
quantitatively in the presence of perturbations. Hence this mode represents a backreaction of the perturbations on the background evolution
\cite{Buchert:1995fz,Buchert:2007ik,Rasanen:2011ki,Buchert:2011sx},
and if it becomes too large, the background model will be a poor description of the true dynamics and has to be improved.

\subsection{The particle ensemble}
\label{ssec:particles}

In order to study the evolution of structure in the Universe we require a description of all the relevant sources
of stress-energy. In this section we are concerned with the description of free-streaming particles,
such as cold dark matter or massive neutrinos, in terms of an ensemble of $N$-body particles. Other sources of stress-energy
will be discussed in the next section.

The cold dark matter paradigm states that a large proportion of the clustering mass in our Universe can
be modelled to good accuracy by a species of non-relativistic (cold), non-interacting (dark) massive
particles. We shall therefore consider an ensemble of classical point-particles as a source of
stress-energy.  At first we make no assumption about the velocity distribution such that this
picture can also be used for particles with relativistic velocities, e.g.\  neutrinos. The
action of an ensemble of particles is simply the sum over the one-particle actions,
\begin{equation}
 \label{e:ensembleaction}
 \mathcal{S}_\mathrm{m} = -\sum_n m_{\left(n\right)} \int \rmd^3 \mathbf{x} \rmd\tau \delta^{\left(3\right)}\!\left(\mathbf{x} - \mathbf{x}_{\left(n\right)}\right)
 \sqrt{-g_{\mu\nu} \frac{\rmd x_{\left(n\right)}^\mu}{\rmd\tau}\frac{\rmd x_{\left(n\right)}^\nu}{\rmd\tau}} \, .
\end{equation}
Here, $m_{\left(n\right)}$, $\mathbf{x}_{\left(n\right)}$ and $\rmd x_{\left(n\right)}^\mu / \rmd\tau$
denote, respectively, the rest mass, position and coordinate four-velocity of the $n$th particle. The
corresponding stress-energy tensor can be obtained as usual by computing the variation with respect to
$\delta g_{\mu\nu}$:
\begin{equation}
 \label{e:ensembleTmunu}
 T_\mathrm{m}^{\mu\nu} = \sum_n m_{\left(n\right)} \frac{\delta^{\left(3\right)}\!\left(\mathbf{x} - \mathbf{x}_{\left(n\right)}\right)}{\sqrt{-g}} \left(-g_{\alpha\beta}
\frac{\rmd x_{\left(n\right)}^\alpha}{\rmd\tau}\frac{\rmd x_{\left(n\right)}^\beta}{\rmd\tau}\right)^{\!-1/2}
\frac{\rmd x_{\left(n\right)}^\mu}{\rmd\tau}\frac{\rmd x_{\left(n\right)}^\nu}{\rmd\tau} \, .
\end{equation}
Expanding to first order in the metric perturbations, this becomes
\begin{eqnarray}
 \label{e:ensembleTmunu2}
 \fl T_\mathrm{m}^{\mu\nu} = \sum_n \frac{m_{\left(n\right)}}{a^5} \delta^{\left(3\right)}\!\left(\mathbf{x} - \mathbf{x}_{\left(n\right)}\right) \gamma_{\left(n\right)} \frac{\rmd x_{\left(n\right)}^\mu}{\rmd\tau}\frac{\rmd x_{\left(n\right)}^\nu}{\rmd\tau}
\left[1 - \Psi + 3 \Phi\phantom{\left(\frac{\rmd x_{\left(n\right)}^j}{\rmd\tau}\right)}\right.\nonumber\\
\left. - \gamma_{\left(n\right)}^2 \left(B_i \frac{\rmd x_{\left(n\right)}^i}{\rmd\tau} - \frac{1}{2} h_{ij} \frac{\rmd x_{\left(n\right)}^i}{\rmd\tau} \frac{\rmd x_{\left(n\right)}^j}{\rmd\tau} + \Psi + \Phi \delta_{ij} \frac{\rmd x_{\left(n\right)}^i}{\rmd\tau} \frac{\rmd x_{\left(n\right)}^j}{\rmd\tau}\right)\right]\, ,
\end{eqnarray}
where $ \gamma_{\left(n\right)}$ is the unperturbed Lorentz factor given by
\begin{equation}
 \gamma_{\left(n\right)} \doteq \left(1 - \delta_{ij} \frac{\rmd x_{\left(n\right)}^i}{\rmd\tau} \frac{\rmd x_{\left(n\right)}^j}{\rmd\tau}\right)^{\!-1/2} \, .
\end{equation}

The evolution of the particle ensemble is obtained by integrating the geodesic equation for each particle,
\begin{equation}
 \frac{\rmd^2 x_{\left(n\right)}^\mu}{\rmd s^2} + \Gamma^\mu_{\alpha\beta} \frac{\rmd x_{\left(n\right)}^\alpha}{\rmd s} \frac{\rmd x_{\left(n\right)}^\beta}{\rmd s} = 0 \, .
\end{equation}
Expanding the Christoffel symbols to first order in the metric perturbations and eliminating the proper time
parameter $s$ in favour of coordinate time $\tau$, the geodesic equation reads
\begin{eqnarray}
\label{e:fullgeodesic}
 \fl\frac{\rmd^2 x_{\left(n\right)}^i}{\rmd\tau^2} + \left(\mathcal{H} - \Psi' - 2 \Phi'\right) \frac{\rmd x_{\left(n\right)}^i}{\rmd\tau} + \delta^{ij} \left(\Psi_{,j} - \mathcal{H} B_j - B_j' - 2 B_{\left[j,k\right]} \frac{\rmd x_{\left(n\right)}^k}{\rmd\tau} + h_{jk}' \frac{\rmd x_{\left(n\right)}^k}{\rmd\tau}\right) \nonumber\\
 - 2 \left(\Psi_{,j} + \Phi_{,j} - \mathcal{H} B_j\right) \frac{\rmd x_{\left(n\right)}^i}{\rmd\tau} \frac{\rmd x_{\left(n\right)}^j}{\rmd\tau} + \delta^{il} \left(h_{l\left(j,k\right)} - \frac{1}{2} h_{jk,l}\right)  \frac{\rmd x_{\left(n\right)}^j}{\rmd\tau} \frac{\rmd x_{\left(n\right)}^k}{\rmd\tau} \nonumber\\
 + \delta^{il} \delta_{jk} \left(\mathcal{H} B_l + \Phi_{,l}\right) \frac{\rmd x_{\left(n\right)}^j}{\rmd\tau} \frac{\rmd x_{\left(n\right)}^k}{\rmd\tau} - \left(\mathcal{H} h_{jk} + \frac{1}{2} h_{jk}' + B_{\left(j,k\right)}\right) \frac{\rmd x_{\left(n\right)}^i}{\rmd\tau} \frac{\rmd x_{\left(n\right)}^j}{\rmd\tau} \frac{\rmd x_{\left(n\right)}^k}{\rmd\tau}\nonumber\\
 - \delta_{jk} \left(\mathcal{H} - 2 \mathcal{H} \Psi - 2 \mathcal{H} \Phi - \Phi'\right) \frac{\rmd x_{\left(n\right)}^i}{\rmd\tau} \frac{\rmd x_{\left(n\right)}^j}{\rmd\tau} \frac{\rmd x_{\left(n\right)}^k}{\rmd\tau} = 0 \, ,
\end{eqnarray}
where we have introduced the conformal Hubble parameter $\mathcal{H} \doteq a'/a$. To linear order in
velocities and in the metric perturbations, this expression simplifies to
\begin{equation}
 \label{e:geodesic}
 \frac{\rmd^2 x_{\left(n\right)}^i}{\rmd\tau^2} + \mathcal{H} \frac{\rmd x_{\left(n\right)}^i}{\rmd\tau} + \delta^{ij} \left(\Psi_{,j} - \mathcal{H} B_j - B_j'\right) = 0 \, .
\end{equation}
Hence, nonrelativistic particles are only sensitive to
the gradient of $\Psi$ and to frame dragging, but not to $\Phi$ or gravitational waves, while the propagation of
relativistic particles is affected by all types of metric perturbations.

For ultrarelativistic particles, integrating the geodesic equation in terms of
$\rmd x_{\left(n\right)}^i / \rmd\tau$ is dangerous, because numerical errors can easily take the
particle trajectory out of the light cone. In this situation, it is advised to work with a different
variable which is not bounded from above by causality.
Possibilities include using either
$\gamma_{\left(n\right)} \left(\rmd x_{\left(n\right)}^i / \rmd\tau\right)$ or $a u^i\doteq a \left(\rmd x_{\left(n\right)}^i / \rmd s\right)$ as a new variable;
$u^0 \doteq \left(\rmd x_{\left(n\right)}^0 / \rmd s\right)$ is then determined by the normalization condition $g_{\mu\nu} u^\mu u^\nu=-1$.

\subsection{Additional sources}

Apart from free-streaming particles, the late Universe contains also other sources of stress-energy, most notably dark energy.
Probably the simplest example of  dark energy is a cosmological constant which is still viable~\cite{Ade:2013zuv} and has the stress-energy tensor
$T^{\mu}_{(\Lambda)\nu} = -(\Lambda / 8 \pi G) \delta^{\mu}_{\nu}$. Since it contains no perturbations -- its
energy density is the same in any frame and at any position -- it appears explicitly only in the
background equations. However, the time dependence of the background has an impact on the evolution
of the perturbations. A positive $\Lambda$ leads to a higher expansion rate and therefore to
stronger Hubble damping.

Since the true nature of dark energy is unknown, it is interesting to consider also
alternatives. Models where dark energy is generated by dynamical degrees of freedom typically would
contain perturbations whose effect may give useful hints about the nature of these degrees of
freedom and their interactions \cite{Amendola:2012ys}. Given that a better understanding of dark energy constitutes one
of the main challenges of cosmology for the coming years, it is important to be able
to simulate structure formation for various models of dynamical dark energy. Such simulations lead to
predictions for these models in the nonlinear regime which are not accessible to analytical
calculations, and can therefore extend significantly the leverage for observational model inference.

In principle, once an action is specified, one can write down the equations of motion and the
stress-energy tensor for any model. Unfortunately, this does not mean that any model can be solved
numerically in practice. This question depends on the structure of the equations and on the
ability to either resolve the relevant scales or to find reasonable approximation methods. We
cannot give a general recipe here -- different classes of models have to be considered on a
case-by-case basis.

There are some sources of stress-energy for which a numerical treatment has already been
worked out successfully in the past, for example topological defects \cite{Obradovic:2011mt} or several
classes of modified-gravity models \cite{Oyaizu:2008sr,Chan:2009ew,Li:2011vk,Puchwein:2013lza}.
Another important example in cosmology are baryons, for which
different hydrodynamical schemes have been developed in the context of Newtonian mechanics
\cite{Colella:1982ee,Gingold:1977sh}. Since the velocities of baryons are small, it will be
straightforward to generalize these schemes such that they fit into our relativistic framework.
Essentially, one simply needs to extract the stress-energy tensor (or a good approximation
to it) from the hydrodynamical variables and use a relativistic prescription for the gravitational
acceleration, e.g.\ given by \eref{e:geodesic}, with a pressure force term on the right hand side. 
If one wants to go beyond that, one can also consider
 accommodating electromagnetic fields and their interactions with the baryon plasma, in
other words, magneto-hydrodynamics. Such interesting applications are, however,  beyond the
scope of this article. Here we assume that on sufficiently large scales hydrodynamical processes can
be neglected and baryons can simply be considered as part of the dark matter. In this case one can still
include baryonic features in the initial conditions \cite{Eisenstein:1997ik}, which has become a standard procedure for
doing purely gravitational $N$-body simulations for cosmology. The effect of bias, i.e.\ the difference
in clustering of dark matter and baryons on small scales where baryons form galaxies, can be taken into account by
`halo finder' and `halo occupation' algorithms \cite{Berlind:2001xk}.

\section{Einstein's equations}\label{sec:einstein}

The metric variables of \eref{e:metric} are determined by Einstein's field equations. We want
to find solutions in a cosmological context and we argued above that under this premise the metric perturbation variables
are expected to be small. Therefore, we employ an expansion in terms of these variables in order
to obtain a more tractable set of equations -- solving Einstein's equations in full generality remains too
grand a challenge, even 100 years after their discovery. There are different approaches of how to perform such an
expansion systematically, and they lead to varying choices about which terms to keep at a given order. For instance,
a straightforward Taylor expansion leads to very complicated equations already at second order, but not all
the second-order terms need to be equally relevant given a physical situation.

Let us recall the situation we want to address. We want to describe the Universe on large scales, i.e.\
above the scale of compact objects, but we do not want to be limited to scales within the horizon. Furthermore, we
do not want to place strong constraints on the stress-energy tensor, since we may want to study some exotic
models beyond $\Lambda$CDM. We therefore choose a particular scheme that
was introduced in
\cite{Green:2011wc,Adamek:2013wja}, and which is well suited for this purpose. At leading order, it contains both the small-scale Newtonian limit and
large-scale relativistic linear perturbation theory. At second order, so-called ``shortwave corrections'' are introduced which take
into account the most relevant effects of non-linear clustering on small scales, see \sref{sec:shortwave}. 
Since the scheme is in principle valid on all scales with the exception of high-curvature regions near black holes as
mentioned in assumption (\ref{a:smooth}),
we are able to recognize when our assumptions break down, i.e.\ when the metric perturbations become large. We can therefore be
confident that the results represent the general-relativistic dynamics correctly as long as they do not leave the range of
validity of our framework. Let us further stress that
this scheme never assumes the Newtonian approximation to be valid, which means that it can handle relativistic stress-energy
sources as long as they lead to metric fluctuations which remain small.\footnote{The shortwave approximation was originally
developed in a Newtonian context \cite{Green:2011wc,Adamek:2013wja} where it was appropriate to include a low velocity approximation.
However, this is not an intrinsic feature of the scheme and can easily be relaxed.}

In this respect, the scheme follows a slightly different philosophy than a post-Newtonian expansion. Such an
expansion was developed and successfully applied in a cosmological context in \cite{Milillo2010,Bruni:2013mua}.
There the idea is to restore $c$ in all equations and employ an expansion in inverse powers of $c$. As one may expect, the leading order
of this expansion \textit{is} the Newtonian limit. What is interesting and reassuring is 
that in a context where an expansion of the stress-energy tensor in terms of inverse powers of $c$ is appropriate, in particular
for cold dark matter,
also the higher-order terms in the post-Newtonian expansion have a direct counterpart in our relativistic shortwave expansion.

At this point it may be useful to establish the correspondence in more detail.
Instead of $(1 + 2 \Psi)$ and $(1 - 2 \Phi)$ in \eref{e:metric} let us
write $e^{2 \tilde{\Psi}}$ and $e^{-2 \tilde{\Phi}}$, respectively.\footnote{To some readers it may have been more appealing
to use $\tilde{\Psi}$ and $\tilde{\Phi}$ in the first place. However, as long as $|\Phi|, |\Psi| \ll 1$ in accordance with assumption
(\ref{a:pert}), their relation can always be inverted, and we choose to use $\Psi$ and $\Phi$ to conform with conventions previously
used in literature \cite{Green:2011wc,Adamek:2013wja}.} Then, one can formally expand $\tilde{\Psi}$ as
\begin{equation}
 \tilde{\Psi} = -\frac{U_N}{c^2} - \frac{2 U_P}{c^4} + \ldots \, ,
\end{equation}
and similarly for $\tilde{\Phi}$. Here, $U_N$ is the dimensionful Newtonian potential, $U_P$ is the first post-Newtonian correction, and
the notation and numerical coefficients follow \cite{Bruni:2013mua}. For the metric perturbations
$B_i$ and $h_{ij}$ one also employs such expansions,
but starting with $c^{-3}$ and $c^{-4}$, respectively, as the order of the leading terms, cf.\ \cite{Milillo2010,Bruni:2013mua}.
Finally, one writes $\rmd x_{\left(n\right)}^i / \rmd\tau = v_{\left(n\right)}^i / c$, which means that now both sides of
Einstein's equations can be expanded in inverse powers of $c$. One then solves order-by-order for the coefficients in the expansions.
From this procedure
it is evident that any term in either expansion also appears in the other, albeit possibly at a different order.

In our view, the main drawback of the post-Newtonian approach is its intrinsic limitation to non-relativistic
sources. For instance, the expansion does not formally converge for a contribution from a relativistic particle species
(e.g.\ neutrinos), even though they may actually only give minor corrections to the metric perturbations. Our approach
does not rely on such a strong assumption about the stress-energy tensor which is why we think that it is more robust
and flexible, also with regard to applications to models beyond $\Lambda$CDM.

In the following sections we explain our scheme in more detail. We also point out the correspondence
to the post-Newtonian framework at appropriate instances.

\subsection{The lowest order}
\label{sec:1storder}

To lowest order in  metric perturbations, the time-time component of Einstein's equations reads
\begin{equation}
 \label{e:G00}
 \Delta \Phi - 3 \mathcal{H} \Phi' - 3 \mathcal{H}^2 \Psi = -4 \pi G a^2 \delta T_0^0 \, ,
\end{equation}
where $\delta T_0^0 \doteq T_0^0 - \bar{T}_0^0$, and $\bar{T}_0^0$ is the model for the background
stress-energy tensor that is used to solve Friedmann's equations for the scale factor $a$. As explained earlier, the
precise prescription for how to construct $\bar{T}^{\mu\nu}$ does not really matter because the
difference between different prescriptions will be accounted for by the homogeneous mode of $\Phi$.
This means that any prescription is good as long as this homogeneous mode does not become  large so
that the framework breaks down. We also stress again that we do not assume $\delta T_0^0$ to be small.
Otherwise we would simply be doing relativistic perturbation theory.

As is well known, not all of Einstein's equations are needed to find the metric perturbations from
given stress-energy perturbations.
Together with the time-time equation, it is sufficient to consider the
traceless part of the space-space components of Einstein's equations.
To lowest order in the metric perturbations, these are given by 
\begin{eqnarray}
\label{e:Gij}
 \fl \qquad \frac{1}{2} h_{ij}'' + \mathcal{H} h_{ij}' - \frac{1}{2} \Delta h_{ij} + B_{\left(i,j\right)}' + 2 \mathcal{H} B_{\left(i,j\right)}
 + \left(\Phi - \Psi\right)_{,ij} - \frac{1}{3} \delta_{ij} \Delta \left(\Phi - \Psi\right) \nonumber\\
 = 8 \pi G a^2 \left(\delta_{ik} T^k_j - \frac{1}{3} \delta_{ij} T_k^k\right) \doteq 8 \pi G a^2 \Pi_{ij}\, .
\end{eqnarray}
Here $\Pi_{ij}$ is the anisotropic stress.
Decomposing this tensor equation into spin-0, spin-1, and spin-2 components,
one obtains equations for $\left(\Phi - \Psi\right)$, $B_i$, and $h_{ij}$, respectively.
This decomposition is most conveniently done in Fourier space where it is given by local projection
operators.\footnote{We have adopted a new
strategy here compared to the previous work \cite{Adamek:2013wja} where the second scalar and the vector were supposed to
be extracted from the spatial trace and the time-space components of Einstein's equations, respectively. Our new strategy
has the advantage that it requires less operations, less memory, and fewer code components. The disadvantage is that the
algorithm we propose, being based on the Fourier method, requires a structured lattice and can not easily be
generalized to a framework with adaptive mesh refinement.}

Alternatively, one can extract $B_i$ from the time-space components of Einstein's equations, which are
\begin{equation}
 \label{e:G0i}
 -\frac{1}{4}\Delta B_i - \Phi_{,i}' - \mathcal{H} \Psi_{,i} = 4 \pi G a^2 T^0_i\, .
\end{equation}
Their transverse projection gives a constraint from which $B_i$ can be obtained directly.

\subsection{The Newtonian limit}

If the anisotropic stress can be neglected, which is a fairly good approximation if $T^{\mu\nu}$ is
dominated by nonrelativistic matter, it follows from \eref{e:Gij} that it is consistent to neglect vector and tensor modes, and that the
two scalar potentials $\Phi$ and $\Psi$ are identical. In this case, one needs to solve only one scalar equation,
which is obtained from \eref{e:G00} by setting $\Psi = \Phi$.

Considering equation \eref{e:G00} in Fourier space, one observes that the second and third term on the left-hand side can only compete
with the first term on scales $k \lesssim \mathcal{H}$. On length scales which are well inside the horizon, the first term
 dominates and one may neglect the other terms, which yields
\begin{equation}
 \label{e:Newton}
 \Delta \Phi = -4 \pi G a^2 \delta T_0^0 \, . \qquad \mathrm{(sub~horizon)}
\end{equation}
This corresponds to the equation for the Newtonian potential, once we identify $\delta T_0^0$ with the density
perturbation. Again we see that General Relativity is well approximated by Newtonian gravity in the limit where
velocities and potentials are small, and the scales of interest are well inside the Hubble horizon. As noted previously,
the Newtonian limit is also obtained naturally  at leading order when expanding in inverse powers of $c$ (after restoring $c$ in the equations).
From a post-Newtonian perspective the effect of a finite horizon and, for instance, the effect of
relativistic kinetic energy both appear at next-to-leading order. A thorough discussion of the validity of the Newtonian approximation
in cosmology can be found in the contribution by Green and Wald to this focus issue, or in earlier references, e.g.\ \cite{Green:2011wc,Chisari:2011iq}.

\subsection{Shortwave corrections}
\label{sec:shortwave}

A systematic way to improve on the lowest order description presented in \sref{sec:1storder} can be found by following
considerations. As outlined in assumption (\ref{a:shortwave}) we expect that, even though metric perturbations remain small on all 
cosmological scales, they can exhibit short wavelength fluctuations which give rise to large spatial derivatives and/or large fluctuations
of the curvature. Note that this does not happen
in the radiation dominated regime where metric perturbations decay once they enter the horizon. In the matter dominated regime,
on the other hand, scalar perturbations of the metric remain approximately constant at all scales and can
hence lead to large
spatial derivatives. Instead of including all quadratic
terms it therefore makes sense to include only those terms which are enhanced by spatial derivatives. We call such terms ``shortwave corrections'' because they
account for the most important second-order terms in the presence of short wavelength fluctuations.

Technically, in order to obtain all these terms in a systematic way, one can simply give every spatial derivative a weight
$\epsilon^{-1/2}$, where $\epsilon$ is an expansion parameter that characterizes the smallness of metric perturbations.
One then retains all terms up to order $\epsilon$. In other words, a term which is quadratic in the metric perturbation
variables is retained if and only if it comes with two spatial derivatives. More details about this procedure can be found
in \cite{Green:2011wc,Adamek:2013wja}. Note that we do not include quadratic terms from time derivatives
of metric perturbations. This is justified by the fact that relativistic perturbations, which
may lead to fast oscillations, decay inside the horizon and can therefore be treated accurately within linear perturbation theory.

It turns out that the second order shortwave corrections obtained this way are all constructed from the two scalars $\Phi$
and $\Psi$. In particular, whereas \eref{e:G0i} remains unchanged, \eref{e:G00} becomes
\begin{equation}
\label{e:G00sw}
 \left(1 + 4 \Phi\right) \Delta \Phi - 3 \mathcal{H} \Phi' - 3 \mathcal{H}^2 \Psi + \frac{3}{2} \delta^{ij} \Phi_{,i} \Phi_{,j} = -4 \pi G a^2 \delta T_0^0 \, ,
\end{equation}
and \eref{e:Gij} turns into
\begin{eqnarray}
\label{e:Gijsw}
 \fl \frac{1}{2} h_{ij}'' + \mathcal{H} h_{ij}' - \frac{1}{2} \Delta h_{ij} + B_{\left(i,j\right)}' + 2 \mathcal{H} B_{\left(i,j\right)}
 + \left(\!\frac{\partial^2}{\partial x^i \partial x^j} - \frac{1}{3} \delta_{ij} \Delta\!\right) \!\left[\left(\Phi - \Psi\right) \left(1 + \Phi - \Psi\right) + \Phi^2\right] \nonumber\\
 + 2 \Psi \Phi_{,ij} - \frac{2}{3} \delta_{ij} \Psi \Delta \Phi - \left(\Phi - \Psi\right)_{,i} \left(\Phi - \Psi\right)_{,j} + \frac{1}{3} \delta_{ij}
 \delta^{kl} \left(\Phi - \Psi\right)_{,k} \left(\Phi - \Psi\right)_{,l}\nonumber\\
 = 8 \pi G a^2 \Pi_{ij}\, .
\end{eqnarray}
While the first line still separates neatly into spin-2, spin-1, and (nonlinear) spin-0 terms, the second line now contains mixed terms
which complicate the treatment of the equation. We will discuss strategies for the numerical solution of \eref{e:G00sw} and \eref{e:Gijsw}
in \sref{sec:numerics}.

The stress-energy tensor may depend on the metric, but it does not contain any derivatives thereof, and therefore
does not acquire any shortwave corrections. However, if one also employs an expansion in low velocities, like in the post-Newtonian framework, corrections quadratic in velocities
should also be taken into account at this level. In our framework this can be understood by noting
that at leading order velocities are proportional to the gradient
of $\Psi$ -- see \eref{e:geodesic} -- and we may formally assign to them an expansion weight $\epsilon^{1/2}$. This means,
for instance, that the anisotropic stress of cold dark matter is expected to be of the same order as the shortwave corrections.

Expanding the geodesic equation to order $\epsilon$ in this counting scheme yields one single correction term, such that
the geodesic equation at low velocities becomes\footnote{This corrects a sign error in \cite{Adamek:2013wja}.}
\begin{equation}
 \label{e:geodesicsw}
 \frac{\rmd^2 x_{\left(n\right)}^i}{\rmd\tau^2} + \mathcal{H} \frac{\rmd x_{\left(n\right)}^i}{\rmd\tau} + \delta^{ij} \left(\Psi_{,j} - \mathcal{H} B_j - B_j' - 2 B_{\left[j,k\right]} \frac{\rmd x^k_{\left(n\right)}}{\rmd\tau}\right) = 0 \, .
\end{equation}
It should be stressed that this approximation is valid only if the individual particle velocities are small. Even if
the particles are relativistic,
the energy flow velocity field $T_0^i$ can be small since it is the average of all particle velocities
in one volume element. For this reason the smallness of $T_0^i$ is 
not the relevant criterion for an expansion in low velocities.
If necessary, one should use the full geodesic equation~\eref{e:fullgeodesic}.

\subsection{Initial conditions}

In order to set up a simulation of the Universe, we need to specify Cauchy data for the metric, the particle ensemble, and
all other dynamical degrees of freedom which constitute additional sources in the model of interest, on a spacelike hypersurface.
This initial data is determined by the physical processes that take place in the early Universe before the initial time of the simulation.
In the case where these processes can be described by linear perturbation equations one can use  a linear Boltzmann solver like
CAMB~\cite{Lewis:1999bs} or CLASS~\cite{Blas:2011rf} to obtain the initial data.
 A numerical
simulation is typically initialized just before linear perturbation theory  is about to break down
 because some perturbation variable is becoming large. The relativistic framework presented here remains
valid beyond this point since it is fully nonperturbative in the stress-energy tensor. We employ a perturbative expansion in the metric variables
which remain small throughout, but the evolution of matter is nonperturbative.

For many cosmological scenarios, in particular $\Lambda$CDM standard cosmology and its natural extensions (massive neutrinos, additional
species etc.), a perturbative treatment has long been established. Initial conditions for a relativistic simulation can therefore
be calculated in these scenarios and just need to be translated into the longitudinal gauge. Details on how this is done in
the context of linear perturbation theory
of $\Lambda$CDM are given in appendix B of \cite{Adamek:2013wja} and many textbooks on cosmology, e.g.\ \cite{1998ARA&A..36..599B,2003:Aar}.
The use of higher-order perturbation theory results is a bit more
cumbersome, but it would allow to include the shortwave corrections already at the level of initial data. If all perturbations
are small, it is consistent to set these quadratic terms to zero initially. Their amplitude is then one of the limiting factors
of the accuracy of the scheme (see e.g.\ \cite{Crocce:2006ve,Tatekawa:2007ix} for a discussion).

\section{Numerical recipes}
\label{sec:numerics}

In reality, we can imagine taking the continuum limit where the number of particles becomes very
large and the individual particle masses become very small. For instance, if cold dark matter is made up of
particles with a mass in the TeV range, the average number density in the present-day Universe is of the order of $10^{46}$
particles per cubic parsec. Since the numerical difficulty grows at least proportional to the number
of particles, simulations can only handle particle ensembles up to a certain size which is typically
many orders of magnitude smaller than the physical ensemble. The presently largest $N$-body 
simulations~\cite{Kim:2011ab,Ishiyama:2012gs} comprise about $10^{12}$ particles. If we would want these
to represent elementary particles we could only describe a volume of about $\sim (100$~km$)^3$. Therefore we have to
choose much larger particle masses. It is more appropriate to consider these as discrete elements of the physical
six-dimensional phase space than as individual particles. The precise value of the mass depends on both the volume we want
to simulate and the computational resources at our disposal, i.e.\ the resolution we can reach.
Similarly, while the metric is a continuous
field, a numerical treatment requires the field to be discretized, for instance by tessellating
the coordinate volume into a cubic lattice. The field equations are then approximately represented as
finite-difference equations. In the following sections, we explain in some detail how one can
numerically solve the coupled evolution of the fields on a lattice and the particle ensemble.

An effective phase space description like the one adopted by N-body simulations may appear more cumbersome
than the fluid description usually invoked in perturbation theory. This additional level of complexity is, however,
unavoidable since the fluid approximation breaks down during nonlinear evolution. A fluid description is only appropriate
until the free streaming dark matter particles encounter shell crossing, i.e.\ until two streams of particles with different
velocities meet in the same volume element~\cite{Adamek:2014qja}.
Clusters are composed of many overlapping streams where the phase space distribution of dark matter particles
is highly non-trivial and cannot be represented by the lowest-order fluid variables of the Boltzmann hierarchy.

\subsection{Particle-to-mesh projection and force interpolation}\label{ssec:particle-mesh}

The discretization of the equations makes it necessary to establish a prescription of how to obtain a lattice-based
stress-energy tensor from a particle ensemble. Such a prescription is called ``particle-to-mesh projection.''
On the other hand, in order to solve the evolution of the ensemble using \eref{e:fullgeodesic}, we have to
interpolate the metric field from the lattice to the particle positions.

\begin{figure}[tb]
 \includegraphics[width=0.85\textwidth]{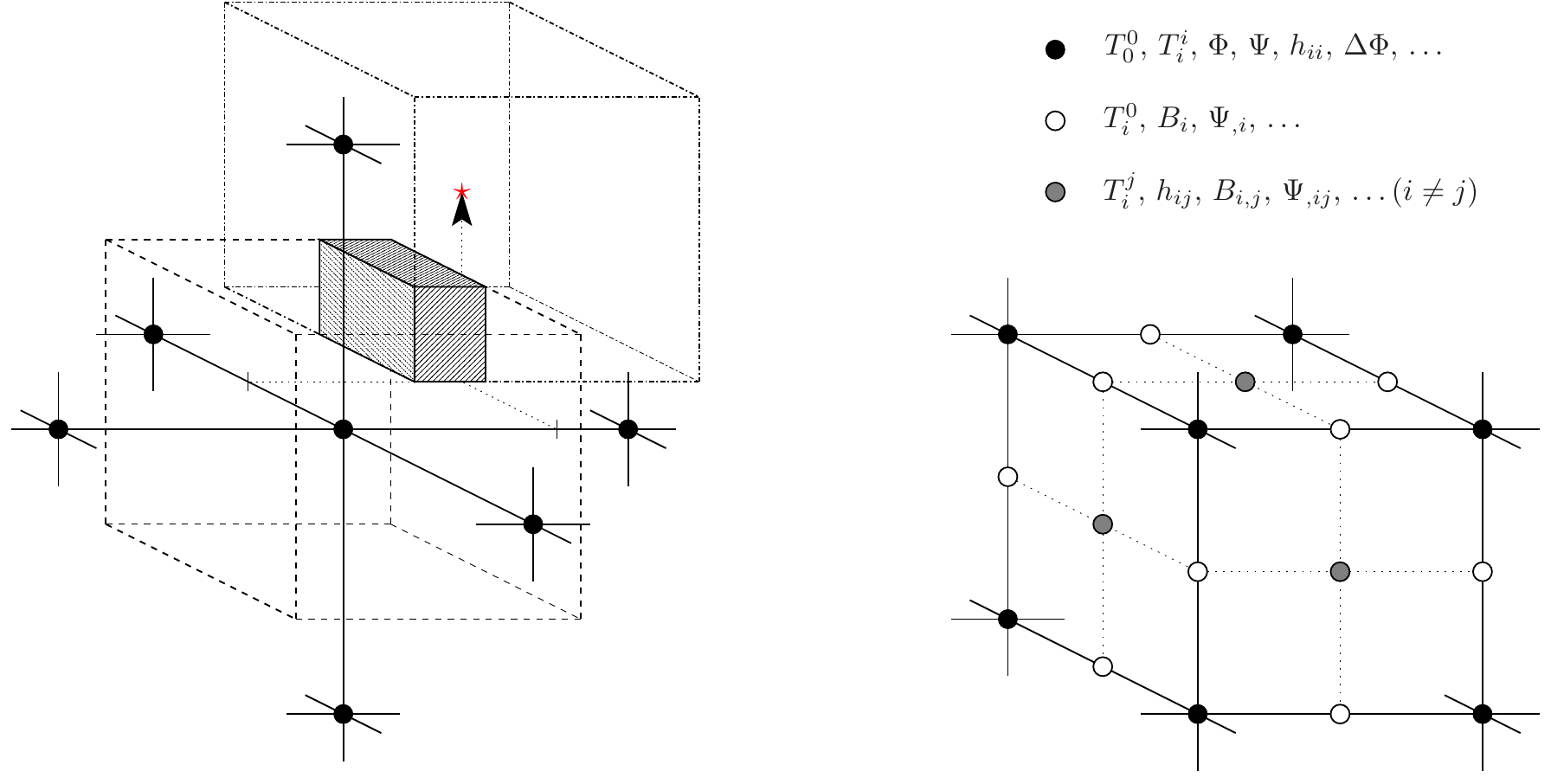}
\caption{\label{fig:CIC} \small Left: Geometric construction of the weight function $w$ which defines the cloud-in-cell projection method.
The weight contributing to the central lattice point is given as the volume fraction (shaded region) of the cubic cloud (dot-dashed), one lattice unit in
size and centered at the particle position (red star), which lies inside the lattice point's Voronoi cell (dashed). --- Right: Location of the lattice
representations of various fields relative to the lattice of scalar quantities (black vertices) if the standard one-sided two-point gradient is used as the definition
of the lattice derivative. In this case, vector fields naturally live on the edges (white vertices) connecting the lattice sites of the primary lattice.
The off-diagonals of tensor fields of rank two are found on the faces of the lattice cells (grey vertices).}
\end{figure}

Since the $N$-body ensemble usually cannot represent the full ensemble of fundamental particles, one should think
of it as a ``representative'' sample of the phase space distribution. For instance, one could imagine that each
$N$-body particle is composed of
a small cloud of fundamental particles which follow approximately the same trajectory in phase space. The stress-energy
tensor of the particle ensemble is then obtained by replacing the delta functions in \eref{e:ensembleTmunu} by
slightly extended density functions representing the shape of the particle clouds:
\begin{equation*}
 \delta^{\left(3\right)}\!\left(\mathbf{x} - \mathbf{x}_{\left(n\right)}\right) \rightarrow w\left(\mathbf{x} - \mathbf{x}_{\left(n\right)}\right)
\end{equation*}

It is convenient to choose a shape for $w$
which is adapted to the lattice spacing such that the evaluation of the stress-energy tensor at the lattice points becomes
numerically a simple operation. Standard approaches can be found in textbooks, for instance \cite{HockneyEastwood}, where
a systematic hierarchy of shape functions is developed which each correspond to a different particle-to-mesh
projection method. To give an explicit example, let us briefly sketch the cloud-in-cell (CIC) scheme.

For CIC projection the function $w$ is constructed as follows. Assume that each $N$-body particle represents a cloud that uniformly fills
a cubic volume of one lattice unit in size, centered at the particle position $\mathbf{x}_{\left(n\right)}$. The weight $w\left(\mathbf{x} - \mathbf{x}_{\left(n\right)}\right)$
evaluated at some point $\mathbf{x}$ (e.g.\ a nearby lattice point) is then given by the fraction of the cloud's volume that lies within the fundamental domain
centered at $\mathbf{x}$. For lattice points the associated fundamental domain is simply given by the corresponding (Voronoi) cell of the dual lattice.
The situation is illustrated in the left panel of \fref{fig:CIC}.

From this geometric construction it is evident that $w$ is simply the triple product of piecewise-linear functions in each direction. Therefore, $w$ is
continuous but not continuously differentiable. The next higher-order method in the hierarchy of projection method has a continuous first derivative of $w$ 
but the second derivative is not continuous, and so forth.

A nice feature of these assignment schemes is that each of them comes with an
interpolation method that can be used to obtain the metric at the particle positions.
Namely, whenever a lattice-based quantity needs to be interpolated to an off-lattice position, one can simply make use of the same function $w$ again.
This time, the sum needs to be taken over all contributing lattice points:
\begin{equation}
 Q(\mathbf{y}) \doteq \sum_{\mathbf{x} \in \mathrm{lattice}} w\left(\mathbf{x} - \mathbf{y}\right) Q_\mathrm{lat}(\mathbf{x})\, ,
\end{equation}
where $Q$ can be any quantity that has a lattice-based representation $Q_\mathrm{lat}$.

We finally need to address the question of how to evaluate the metric-dependent terms in \eref{e:ensembleTmunu2} once the delta functions
have been replaced by $w$. The simplest and most straightforward choice is to evaluate them at the projection point. In this case, at second
order in particle velocities, all metric-dependent terms can be factored out, such that the sum over the particles can be carried out independently
of the metric.\footnote{In \cite{Adamek:2013wja} we called the result of this particle sum the ``bare'' quantities, whereas after multiplication with the factor
that takes into account the perturbed metric they became the ``dressed'' physical quantities.} Another choice would be to evaluate the metric-dependent
terms for each particle at their actual position, or even halfway between projection point and particle position. Such a prescription would make the projection operation
more involved without providing any clear benefit. The metric-dependent terms are only small perturbations after all, and we can therefore expect that the difference
between various prescriptions will be totally negligible. In this case the choice can be guided by convenience of implementation.

When going from a continuum description to a lattice-based one a further subtlety emerges which has to do with the definition of the lattice derivative.
For instance, if one uses the standard one-sided two-point gradient, the components of the gradient vector field naturally live on different lattices than the scalar fields:
the $1$-component is associated with a lattice shifted by half a unit in the $x^1$-direction and similarly for the other two dimensions. It is natural to associate
all vector fields to these shifted lattices, which coincide with the edges of the ``primary'' lattice. For tensors of rank two the situation is even more
complicated. The diagonal components live on the primary lattice, whereas the off-diagonals live on the faces of the primary lattice cells, i.e.\ on lattices
shifted by half a unit in the two directions given by the values of the two tensor indices, respectively.

The logic behind this story can easily be understood by applying lattice derivatives to various quantities and checking that objects with the same number of
unsaturated indices all live on the same type of lattice. For instance, the divergence of a vector field is a scalar, and it will live on the same lattice as all other scalars.
As mentioned earlier, everything hinges on the choice of lattice derivative, and different choices will lead to different results. We will use the standard one-sided two-point
gradient in which case the situation is the one just described, summarized in the right panel of \fref{fig:CIC}.

This exposition also has implications for the interpolation scheme. It should be self-evident that the interpolation of a quantity
has to be done based on the appropriate lattice where the quantity is represented.

To summarize, in order to evolve the metric, the stress-energy of the particle ensemble can be evaluated on the lattice
by employing a standard particle-to-mesh projection. Then, in order to evolve the particles, the metric is interpolated to
the particle positions using the interpolation scheme associated with the projection method.

\subsection{The parabolic solver}
\label{sec:parabolic}

In the following two sections, we discuss numerical methods to solve \eref{e:G00sw} and \eref{e:Gijsw} for the metric
perturbations. Let us begin with \eref{e:G00sw}, which is a nonlinear parabolic equation for the scalar perturbation $\Phi$.
A first worry may be that we are dealing with a coupled evolution of the metric perturbations -- the equation also contains $\Psi$
and may even contain other metric perturbations entering on the right-hand side. However, if an algorithm which is first-order in
time is sufficient, it is possible to effectively decouple the equations completely. One should keep in mind that in longitudinal gauge, the spin-0 and
spin-1 components of Einstein's equations are constraints, so we expect first-order in time accuracy to be sufficient  in many
cases. We will later comment on how one can improve the scheme to second-order in time if required.

Let us consider a first-order in time solution. We can then simply use
the known values of $\Psi$ (and if necessary also $B_i$ and $h_{ij}$) at the time $\tau$ in the discretized version of
\eref{e:G00sw}, to obtain $\Phi$ at the time $\tau + \rmd\tau$. The other metric perturbation variables
at $\tau + \rmd\tau$ are then obtained using \eref{e:Gijsw} as outlined in the next section.

A major issue for numerical evolution algorithms is stability. For instance, an explicit update scheme for a linear parabolic
equation is stable only if the time step $\rmd\tau$ satisfies the Courant-Friedrichs-Lewy condition \cite{Courant:1928}. This condition requires
that information may not propagate farther than one lattice unit within the time step. Since gravitational fields communicate
at the speed of light, such a condition puts a severe constraint on our time step which makes a simulation almost
unfeasible. In particular, a $N$-body particle with a peculiar velocity of the order of $\sim 100$ km/s would take some thousand
time steps to move by one lattice unit.

We therefore adopt a method which does not put such a constraint on our time step. For a linear parabolic equation, a fully
implicit update scheme has this property. In three spatial dimensions, to avoid the inversion of a large sparse matrix, such an
update can be constructed using an operator-splitting method known as the Douglas-Rachford (or alternating-direction implicit)
method \cite{Douglas1956}. With this approach, the update requires the inversion of three tridiagonal matrices  which can
each be done in $\mathcal{O} (N)$ operations using the Thomas algorithm, $N$ being the rank of the 
matrix~\cite{NumRec}. The scheme is unconditionally
stable and first-order accurate in time. We slightly modify it in order to take into account the nonlinear shortwave corrections,
but we do not expect this to affect stability in any way.

Lattice equations for the parabolic solver have been proposed in \cite{Adamek:2013wja}. We can actually simplify these equations
considerably if we divide \eref{e:G00sw} by the factor $\left(1 + 4 \Phi\right)$, which yields
\begin{equation}
 \label{e:G00sw2}
 \Delta \Phi - 3 \mathcal{H} \Phi' - 3 \mathcal{H}^2 \Psi + \frac{3}{2} \delta^{ij} \Phi_{,i} \Phi_{,j} = -4 \pi G a^2 \left(1 - 4 \Phi\right)
 \delta T_0^0 \, ,
\end{equation}
up to terms which we neglect in our expansion. If we now choose to evaluate the shortwave corrections, 
as well as all metric perturbations apart from $\Phi$  at the beginning of the time step, the tridiagonal matrices which occur in
the finite-difference equations of the Douglas-Rachford algorithm only depend on background quantities, leading to a considerable simplification
of the scheme. Introducing the convenient notation $f^\mathbf{n}_{\mathbf{i},\mathbf{j},\mathbf{k}} \doteq
f(\tau_\mathbf{n}, x^i_{\mathbf{i},\mathbf{j},\mathbf{k}})$, where $\mathbf{n}$ is a discrete index labelling the time steps
and $\mathbf{i}$, $\mathbf{j}$, $\mathbf{k}$ are discrete indices labelling the lattice points of a three-dimensional cubic lattice,
the finite-difference equations become
\numparts
\begin{eqnarray}
 \fl\frac{\Phi^{\mathbf{n}+\frac{1}{3}}_{\mathbf{i}-1,\mathbf{j},\mathbf{k}} + \Phi^{\mathbf{n}+\frac{1}{3}}_{\mathbf{i}+1,\mathbf{j},\mathbf{k}} - 2 \Phi^{\mathbf{n}+\frac{1}{3}}_{\mathbf{i},\mathbf{j},\mathbf{k}}}{\left(\rmd x^1\right)^2} - 3 \mathcal{H} \frac{\Phi^{\mathbf{n}+\frac{1}{3}}_{\mathbf{i},\mathbf{j},\mathbf{k}}}{\rmd\tau} = - 3 \mathcal{H} \frac{\Phi^\mathbf{n}_{\mathbf{i},\mathbf{j},\mathbf{k}}}{\rmd\tau} -\frac{\Phi^\mathbf{n}_{\mathbf{i},\mathbf{j}-1,\mathbf{k}} + \Phi^\mathbf{n}_{\mathbf{i},\mathbf{j}+1,\mathbf{k}} - 2 \Phi^\mathbf{n}_{\mathbf{i},\mathbf{j},\mathbf{k}}}{\left(\rmd x^2\right)^2} \nonumber\\
\hspace*{-1.5cm} -\frac{\Phi^\mathbf{n}_{\mathbf{i},\mathbf{j},\mathbf{k}-1} + \Phi^\mathbf{n}_{\mathbf{i},\mathbf{j},\mathbf{k}+1} - 2 \Phi^\mathbf{n}_{\mathbf{i},\mathbf{j},\mathbf{k}}}{\left(\rmd x^3\right)^2} + 3 \mathcal{H}^2 \Psi^\mathbf{n}_{\mathbf{i},\mathbf{j},\mathbf{k}} - 4 \pi G a^2 \delta T_0^0(\tau_\mathbf{n}, x^i_{\mathbf{i},\mathbf{j},\mathbf{k}}) \nonumber\\
\hspace*{-1.5cm} 
 -\frac{3}{2}\left(\frac{\left(\Phi^\mathbf{n}_{\mathbf{i}+1,\mathbf{j},\mathbf{k}} - \Phi^\mathbf{n}_{\mathbf{i}-1,\mathbf{j},\mathbf{k}}\right)^2}{4 \left(\rmd x^1\right)^2} + \frac{\left(\Phi^\mathbf{n}_{\mathbf{i},\mathbf{j}+1,\mathbf{k}} - \Phi^\mathbf{n}_{\mathbf{i},\mathbf{j}-1,\mathbf{k}}\right)^2}{4 \left(\rmd x^2\right)^2} + \frac{\left(\Phi^\mathbf{n}_{\mathbf{i},\mathbf{j},\mathbf{k}+1} - \Phi^\mathbf{n}_{\mathbf{i},\mathbf{j},\mathbf{k}-1}\right)^2}{4 \left(\rmd x^3\right)^2}\right) \nonumber\\   
\hspace*{-1.5cm}  + 16 \pi G a^2 \Phi^\mathbf{n}_{\mathbf{i},\mathbf{j},\mathbf{k}} \delta T_0^0(\tau_\mathbf{n}, x^i_{\mathbf{i},\mathbf{j},\mathbf{k}})\, ,
\label{e:1over3}\\
 \nonumber\\
 \fl\frac{\Phi^{\mathbf{n}+\frac{2}{3}}_{\mathbf{i},\mathbf{j}-1,\mathbf{k}} + \Phi^{\mathbf{n}+\frac{2}{3}}_{\mathbf{i},\mathbf{j}+1,\mathbf{k}} - 2 \Phi^{\mathbf{n}+\frac{2}{3}}_{\mathbf{i},\mathbf{j},\mathbf{k}}}{\left(\rmd x^2\right)^2} - 3 \mathcal{H} \frac{\Phi^{\mathbf{n}+\frac{2}{3}}_{\mathbf{i},\mathbf{j},\mathbf{k}}}{\rmd\tau} = \frac{\Phi^\mathbf{n}_{\mathbf{i},\mathbf{j}-1,\mathbf{k}} + \Phi^\mathbf{n}_{\mathbf{i},\mathbf{j}+1,\mathbf{k}} - 2 \Phi^\mathbf{n}_{\mathbf{i},\mathbf{j},\mathbf{k}}}{\left(\rmd x^2\right)^2} - 3 \mathcal{H} \frac{\Phi^{\mathbf{n}+\frac{1}{3}}_{\mathbf{i},\mathbf{j},\mathbf{k}}}{\rmd\tau}\, ,\\
 \fl\frac{\Phi^{\mathbf{n}+1}_{\mathbf{i},\mathbf{j},\mathbf{k}-1} + \Phi^{\mathbf{n}+1}_{\mathbf{i},\mathbf{j},\mathbf{k}+1} - 2 \Phi^{\mathbf{n}+1}_{\mathbf{i},\mathbf{j},\mathbf{k}}}{\left(\rmd x^3\right)^2} - 3 \mathcal{H} \frac{\Phi^{\mathbf{n}+1}_{\mathbf{i},\mathbf{j},\mathbf{k}}}{\rmd\tau} = \frac{\Phi^\mathbf{n}_{\mathbf{i},\mathbf{j},\mathbf{k}-1} + \Phi^\mathbf{n}_{\mathbf{i},\mathbf{j},\mathbf{k}+1} - 2 \Phi^\mathbf{n}_{\mathbf{i},\mathbf{j},\mathbf{k}}}{\left(\rmd x^3\right)^2} - 3 \mathcal{H} \frac{\Phi^{\mathbf{n}+\frac{2}{3}}_{\mathbf{i},\mathbf{j},\mathbf{k}}}{\rmd\tau}\, .
\end{eqnarray}
\endnumparts
The two last lines in Eq.~\eref{e:1over3} contain the nonlinear shortwave corrections of 
\eref{e:G00sw2}.
Here, the two intermediate solutions $\Phi^{\mathbf{n}+\frac{1}{3}}$ and $\Phi^{\mathbf{n}+\frac{2}{3}}$ should not be understood as the
solution at fractional time steps. They are simply auxiliary variables which allow to write the problem as three linear tridiagonal matrix
equations, but only the final result $\Phi^{\mathbf{n}+1}$ is physically meaningful. As mentioned earlier, the use of \eref{e:G00sw2} instead
of \eref{e:G00sw} leads to uniform matrix entries for the tridiagonal problems. In particular if the Thomas algorithm has to be implemented
in parallel, which is the case if the simulation is too large to fit on a single core, this simplification reduces communication
requirements. An efficient parallelization of the Thomas algorithm is presented in~\cite{MattorWH95}.

\subsection{The Fourier method}
\label{sec:fourier}

In this section we discuss the numerical treatment of the traceless space-space components of Einstein's equations which,
according to \eref{e:Gij} or its shortwave-corrected version \eref{e:Gijsw}, determine the five other  degrees of freedom of the metric,
$\left(\Phi - \Psi\right)$, $B_i$ and $h_{ij}$. The latter two contain two degrees of freedom each due to the gauge conditions,
while the traceless condition reduces the number of independent equations in \eref{e:Gijsw} from six to five, which is exactly the number we need.

To solve the first-order version \eref{e:Gij} of the equations, a simple procedure is to Fourier transform it
and project onto the spin-0, spin-1 and spin-2 components. In Fourier space these projections are local operations. Since
the spin-0 component is an elliptic equation, one directly obtains $\left(\Phi - \Psi\right)$,
\begin{equation}\label{e:phi-psi}
 \left(\Phi - \Psi\right) = -12 \pi G a^2 \frac{k^i k^j}{k^4} \Pi_{ij} \, .
\end{equation}
where $ \mathbf{k}$ is a Fourier wave vector, and we use the same symbols for the Fourier representation of a quantity as for the quantity
itself. Here and in the following we just give the continuum equations. On a lattice, the projection operators built with $k^i$ have to
be constructed with some care, but one can find appropriate local projectors such that the gauge conditions are implemented according
to the desired definition of lattice derivatives. Some discussion on how this is done can be found, e.g.\ in \cite{Figueroa:2011ye}.

The spin-1 component gives a collection of linear first-order ordinary
differential equations, one for each Fourier mode and each polarization of $B_i$,
\begin{equation}
\label{e:Bmode}
(a^2B_A)' = -16 \pi G a^4 \frac{i k^j \mathbf{e}^{*i}_A}{k^2} \Pi_{ij} \, . 
\end{equation}
Here, $A = 1, 2$ is a polarization label and $\mathbf{B}=B_1 \mathbf{e}_1 + B_2 \mathbf{e}_2$ has two possible polarizations corresponding to the two directions $\mathbf{e}_1$, $\mathbf{e}_2$ orthogonal
to the wave vector, $\delta_{ij} \mathbf{e}^i_A k^j = 0$, $\delta_{ij} \mathbf{e}^i_A \mathbf{e}^j_B = \delta_{AB}$. 
Instead of $\mathbf{e}_{A,B}$ we can also project onto the helicity eigenstates, $\mathbf{e}_{\pm} =\frac{1}{\sqrt{2}}(\mathbf{e}_1 \pm i\mathbf{e}_2)$ with $\mathbf{B}= B_+\mathbf{e}_+ + B_- \mathbf{e}_-$ and 
$B_+=\mathbf{B}\cdot\mathbf{e}^*_+,~B_-=\mathbf{B}\cdot\mathbf{e}^*_-$. With this notation, the index $A$ in \eref{e:Bmode} can be either $1,2$ or $\pm$.

Since these mode equations originate from constraints,
we expect that the solutions vary slowly in time, and the time stepping used for the source is sufficient also for solving the mode
equations. We propose to solve the equations by keeping a Fourier representation of $B_i$ in memory which is updated using
\eref{e:Bmode}. An inverse transform has to be computed only whenever $B_i$ is needed in real space, for instance to obtain the
frame dragging force on the particles.

Finally, the spin-2 component results in a collection of linear second-order ordinary differential equations for the Fourier modes and
polarizations of $h_{ij}$,
\begin{equation}
\label{e:hmode}
h_{X}'' + 2 \mathcal{H} h_{X}' + k^2 h_{X} = 16 \pi G a^2 e^{ij}_X \Pi_{ij} \, .
\end{equation}
Because $h_{ij}$ is symmetric, traceless and transverse, there are again only two independent combinations of $\mathbf{e}_A$ and $\mathbf{e}_B$ which are either 
$e_d^{ij} =\frac{1}{\sqrt{2}}\left[\mathbf{e}^i_1\mathbf{e}^j_1-\mathbf{e}^i_2\mathbf{e}^j_2\right]$ and 
$e_\times^{ij} =\frac{1}{\sqrt{2}}\left[\mathbf{e}^i_1\mathbf{e}^j_2+\mathbf{e}^i_2\mathbf{e}^j_1\right]$ or the helicity eigenstates,
$e_{(+2)}^{ij}=\mathbf{e}^i_+\mathbf{e}^j_+ =\frac{1}{\sqrt{2}}\left[ e_d^{ij} +ie_\times^{ij}\right]$ and
$e_{(-2)}^{ij}=\mathbf{e}^i_-\mathbf{e}^j_- =\frac{1}{\sqrt{2}}\left[ e_d^{ij} -ie_\times^{ij}\right]$ .

The second-order character of \eref{e:hmode} is related to the fact that these equations are not constraints, but
real dynamical equations.
While we expect the scalar and vector metric perturbations to vary slowly, the short wave tensor fluctuations oscillate with frequency
$\omega=k$.  Tracking the high-frequency free solutions allowed by \eref{e:hmode} poses a problem if one does not
want to increase the number of time steps dramatically. However, once inside the horizon, the amplitude of a free gravitational
wave decays like $1/a^2$ and the driven solution is expected to vary in time not much more rapidly than the source itself. 
We note also that $h_{ij}$ does not feed back into
the evolution of the particle ensemble in the low-velocity limit, as can be seen from \eref{e:geodesicsw}. The resolved low-frequency part of
$h_{ij}$ can be evolved much in the same way as $B_i$ -- keeping the Fourier representation in memory and evolving it with the mode
equation \eref{e:hmode}. If one wants to follow the rapid oscillations of the free gravitational waves, one can use a Green's function method as outlined 
in~\ref{s:gws} to recover the spectrum, instead of slowing down the simulation, at least if the full real-space
information of $h_{ij}$ is not needed. The same procedure can be applied for any other relativistic component with small amplitude but rapid oscillations
as may occur in a field theoretical model of dark energy, like e.g.\ quintessence.

From \eref{e:phi-psi}, \eref{e:Bmode} and \eref{e:hmode} we expect that
\begin{eqnarray}
\label{e:scaling}
\Phi-\Psi \simeq  \frac{Ga^2\Pi}{k^2 }\,, \quad
B_A  \simeq  \frac{Ga^2\Pi}{k \mathcal{H} }\,, \quad
h_X \simeq  \frac{Ga^2\Pi}{k^2 } \,.
\end{eqnarray}
Here $\Pi$ denotes a typical component of $\Pi_{ij}$. Therefore, we expect that vector perturbations are parametrically larger than scalar and
tensor perturbations on sub-horizon scales by a factor $k/ \mathcal{H}$.

The situation is somewhat more complicated once we decide to include shortwave corrections, i.e.\ once we want  to solve \eref{e:Gijsw}.
The problem is caused by the mixed terms on the left-hand side of \eref{e:Gijsw} which contribute to all three spin components.
A similar problem can be caused by metric perturbations which may appear on the right-hand side due to the way the stress-energy tensor
may depend on the metric. However, since these terms are all small corrections only, our strategy is to approximate them with
the previous time step where the metric perturbations are already known. We then simply treat these terms as part of the source
which has to be Fourier transformed, and the projection on the different spin components determines these components in the same way as
before. Since we know that the Bardeen potentials $\Phi$ and $\Psi$ evolve slowly, we expect that this procedure does not significantly
reduce the accuracy of the scheme. If this approximation should for some reason not be good enough, one can consider to iterate on the
solutions. By this we mean that one first obtains approximate solutions in the way we just described, and then re-inserts these solutions to
improve the quadratic correction terms. This procedure can be iterated until the solutions converge, which they always should as long
as the shortwave corrections are indeed small.

It is worth noting that for cold dark matter the shortwave corrections, which project dominantly into the scalar sector, are
expected to be of the same order as the anisotropic stress.
The estimate of \eref{e:scaling} is therefore too simplistic in this case, as we will see in \sref{ssec:sims}. In particular, using \eref{e:phi-psi}
with the dark matter anisotropic stress only, neglecting the shortwave corrections to this equation, gives a result which can be wrong by several
orders of magnitude. This is already evident from perturbation theory.

\subsection{The update step}

Assembling the individual components of the evolution algorithm which we discussed in the previous sections, a complete time step is
given by following sequence:
\begin{itemize}
 \item Update particle velocities using \eref{e:geodesicsw} -- the metric components at the particle positions are determined from the known
 lattice values using the interpolation scheme (e.g.\ CIC).
 \item Update particle positions using the updated velocities -- as usual, this is done in a leapfrog fashion, i.e.\ velocities are associated
 to half-integer time steps, while positions and acceleration, and therefore also the metric field values, are associated to integer time steps.
 \item Update $\Phi$ using the parabolic solver of \sref{sec:parabolic}.
 \item Perform a particle-to-mesh projection (e.g.\ CIC) to obtain the new $T^{\mu\nu}$ at the lattice points.
 \item Use the Fourier method of \sref{sec:fourier} to obtain $\Psi$ (or rather $\Phi-\Psi$), $B_i$ and $h_{ij}$.
If we are interested in the power spectra and their time evolution we can directly obtain them from the Fourier representation of the corresponding variables.
\end{itemize}
This sequence is appropriate for simulations which contain only dark matter as source of perturbations. If other sources of stress-energy
are relevant, one has to include appropriate update steps for these constituents as well.

\subsection{Improving to second-order in time}

The algorithms for the evolution of the metric components presented here are accurate only to first-order in time. This may be good enough for
most applications, but let us nonetheless outline briefly how one can improve the scheme to obtain second-order in time accuracy. For the parabolic
solver, a possible modification which leads to a second-order scheme has been discussed in \cite{Douglas1962a}. The difficulty in our case is that
the ``source terms,'' composed of $\delta T_0^0$, $\Psi$ and the shortwave corrections, have to be evaluated also at $\tau+\rmd\tau$. These
values are not available and/or depend on the solution. One can implement a predictor-corrector scheme to solve this problem. This means that one uses the first-order
in time scheme to ``predict'' the values at $\tau+\rmd\tau$, which are then used in the second-order in time update step to obtain the ``corrected'' solution.
This approach roughly doubles the number of arithmetic operations, but the second-order in time improvement can easily compensate for this investment
by allowing for larger time steps. The same predictor-corrector
scheme, when coupled to the Fourier method, can also be used to obtain second-order in time accuracy on all the other metric components.

\section{Preliminary Results}\label{sec:prelim}

As our implementation of the numerical scheme is still under development at the time this review is written, we can only discuss some
preliminary results. In \cite{Adamek:2013wja,Adamek:2014qja} we presented some results for a plane-symmetric setup which allowed
an effectively 1D implementation to test part of the algorithms. In the next subsection we briefly summarize these results.
Since the Newtonian approximation is expected to work well for the baseline $\Lambda$CDM cosmology, which is also seen in the 1D tests,
we then proceed with a post-Newtonian estimate of the relativistic terms $(\Phi-\Psi)$, $B_i$ and $h_{ij}$ using the output of a 3D cosmological
simulation carried out with a conventional Newtonian $N$-body code. In the
future we will study the accuracy of these results with a full three-dimensional relativistic code.

\subsection{Plane symmetric simulations}

A crucial question to investigate first is whether the metric perturbations do indeed remain small in the situations of interest for
cosmological simulations. One situation of possible concern
are shell crossing events. Shell crossing happens when the phase space sheet of dark matter folds {back} upon itself, i.e.\
when two different phase space populations pass through the same point in coordinate space. As opposed to fluid descriptions, $N$-body schemes have
a priori no problem with this situation, but for a perfectly cold particle species without thermal velocity dispersion shell
crossing does correspond formally to a divergence in the stress-energy tensor. However, as
discussed in \cite{Adamek:2014qja} the particle acceleration remains small (although in principle discontinuous at the caustic). 
In terms of the two metric potentials we found that both the amplitude and the gradients remain small, and that the delta-like
density spike at the caustic merely translates to a corresponding feature in the curvature, namely a kink in $\Phi$ and $\Psi$.

In \cite{Adamek:2013wja}, although we were limited by the planar symmetry of the simulations, we
also conducted a study on the amplitude of $(\Phi-\Psi)$ generated by anisotropic stresses
in a $\Lambda$CDM inspired large-scale structure formation context. 
We found that the results obtained with the relativistic code agree very well with the estimates which can be obtained from
a purely Newtonian simulation.
This motivates the treatment in the next section, where we run 3D Newtonian simulations to estimate the
relativistic corrections in a more realistic $\Lambda$CDM cosmology.

\subsection{Estimates from Newtonian simulations} \label{ssec:sims}

\begin{figure}[tb]
 \includegraphics[width=\textwidth]{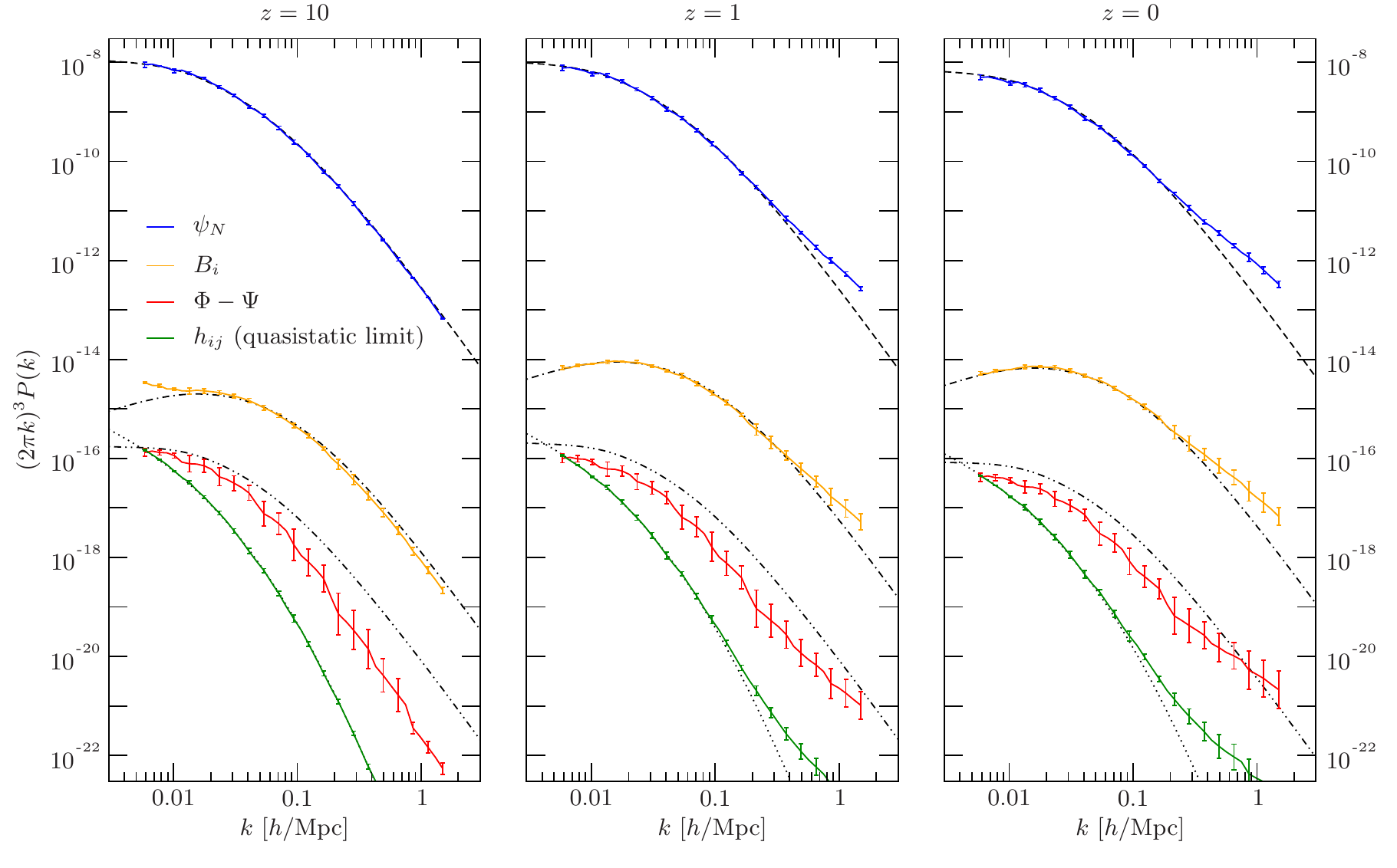}
\caption{\label{fig:3Dspectra} \small Power spectra of the Newtonian gravitational potential $\psi_N$ (blue, highest curve) and of the higher order quantities that vanish
 in linear perturbation theory of $\Lambda$CDM (neglecting radiation): the scalar anisotropic stress $\Phi-\Psi$, the vector perturbation $B_i$ and the
 spin-2 perturbation $h_{ij}$. While the vector power spectrum is about five orders of magnitude smaller than the one of the gravitational potential, the spectra of both,
 $\Phi-\Psi$ and $h_{ij}$, are between seven and ten orders of magnitude smaller.
 The dashed black curve shows the linear power spectrum of the Newtonian potential which was also used to initialize the simulations. We see how at late times and on small scales
 the non-linear evolution of the matter perturbations induces more power in the full spectrum.
The dot-dashed and dotted lines show, respectively, the second order perturbation theory prediction for $B_i$ and $h_{ij}$, based on the linear power spectrum of
the Newtonian potential. The dot-dot-dashed line is the corresponding prediction for $\Phi-\Psi$ using the horizon scale as a cutoff, see text.}
\end{figure}

\begin{figure}[tb]
 \includegraphics[width=\textwidth]{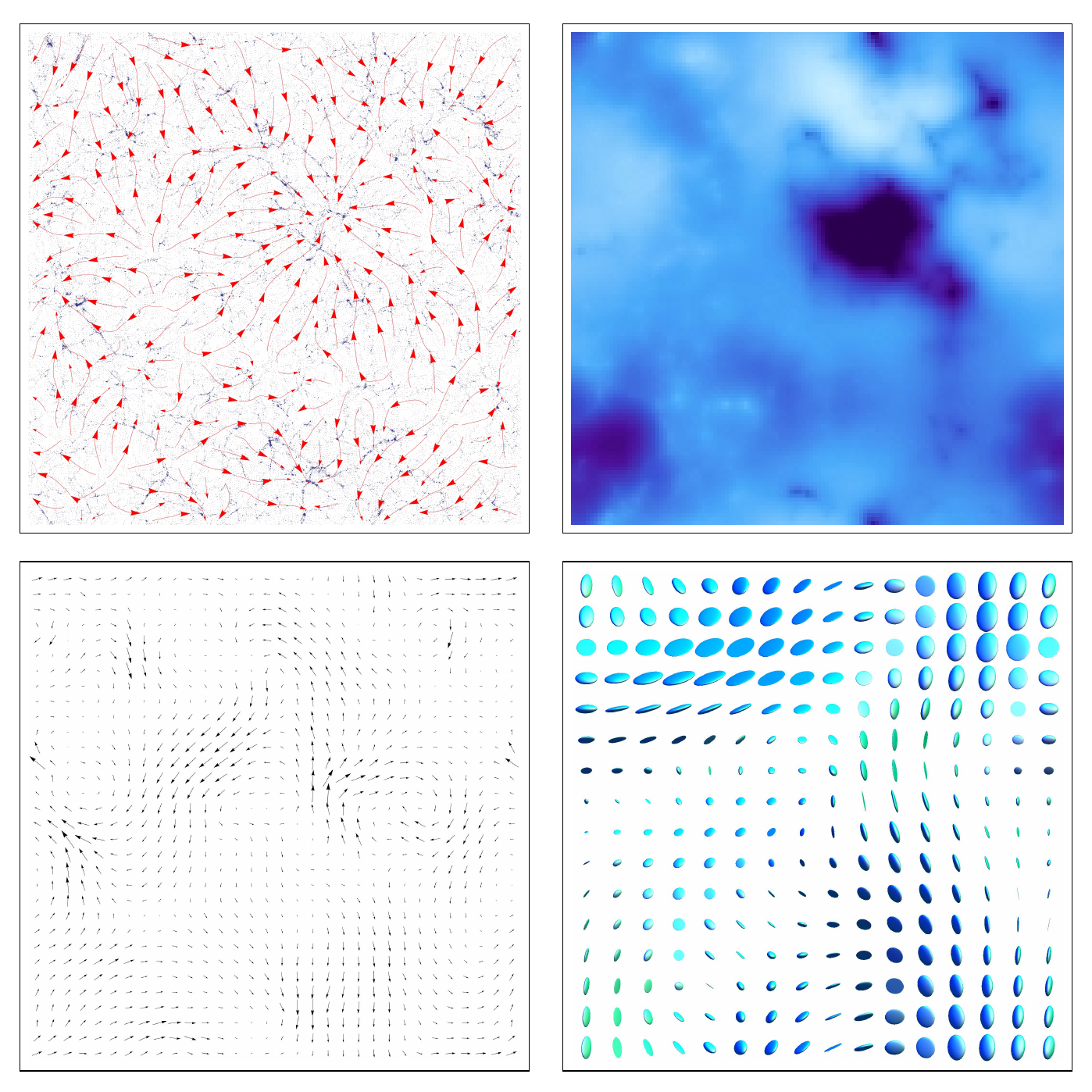}
\caption{\label{fig:slice} \small Cross section through a snapshot taken from a Newtonian $N$-body simulation, showing the real-space reconstructed post-Newtonian metric perturbations.
Top left: Distribution of $N$-body particles and their bulk flow parallel to the figure plane. --- Top right: Map of the scalar post-Newtonian term $\Phi-\Psi$ on the same plane.
--- Bottom left: Vector field showing the two components of the reconstructed spin-1 perturbation $B_i$ which lie within the figure plane. --- Bottom right: Visualization of
the reconstructed spin-2 perturbation $h_{ij}$ along the cross section. The ellipsoids are obtained by applying the local value of $h_{ij}$ as an affine geometric transformation
to some spheroid of fixed radius.}
\end{figure}

In a $\Lambda$CDM cosmology, because velocities are small, the post-Newtonian approach is expected to provide a good approximation.
Thus, in order to obtain estimates of the relativistic corrections in this context,
we proceed as follows. We run a publicly available Newtonian $N$-body code for a $\Lambda$CDM setup and extract the anisotropic stress tensor $\Pi_{ij}$ of the particle ensemble at quadratic
order in the velocities. At leading order we also identify $\Phi$ and $\Psi$ with the Newtonian potential. Then, using \eref{e:Gijsw}, we can obtain an estimate for $\Phi-\Psi$ and $h_{ij}$
at first post-Newtonian order. With this we mean that at that order we can drop the post-Newtonian contributions to the quadratic terms (in particular, we can also drop any squares of $\Phi-\Psi$) and,
furthermore, neglect the time derivatives of $h_{ij}$. The latter approximation amounts to a quasistatic limit of the wave equation for the case where the source term (generated by nonrelativistic matter)
is nearly constant in time. In this case $h_{ij}$ is not a propagating gravitational wave but rather simply the ``static'' spin-2 structure imprinted on the metric due to the presence of anisotropic stresses.

In order to obtain a post-Newtonian estimate for the spin-1 component $B_i$ we use \eref{e:G0i} instead of \eref{e:Bmode}. This allows us
to extract $B_i$ directly from a single snapshot of a simulation without having to integrate
a time-dependent differential equation. It may look appealing to use the same approach within a numerical implementation of the
relativistic scheme, but one should keep in mind that it comes at the cost of three additional particle-to-mesh projections and Fourier
transforms. This is why we anticipate that for the relativistic numerical simulations solving the differential equation may be favourable.

\setcounter{footnote}{0}
The same prescription for estimating the relativistic corrections to the metric has already been explained in great detail in \cite{Green:2011wc}, where (3.14), (3.17) and (3.24) are
essentially the equations we need.\footnote{Equation (3.24) of \cite{Green:2011wc}
contains a sign error on the matter anisotropic stress. Furthermore, all Einstein's equations containing $h_{ij}$ or $\jmath_{ij}$ are missing a factor of $\frac{1}{2}$ on these variables. We introduced the
same mistake in our equation (15) of \cite{Adamek:2013wja}.} Similar equations are derived in \cite{Milillo2010} in the post-Newtonian framework. Note that both references treat matter as a fluid.
We do not use this approximation and extract the anisotropic stress from the full phase space distribution of $N$-body particles as explained in \sref{ssec:particles}.

It should be noted that the procedure outlined above is not entirely self-consistent because  post-Newtonian corrections to the dynamics are not taken into account by the
Newtonian $N$-body code. For the time being we
have to hope that these corrections are irrelevant for our simple estimates  -- the results are at least in qualitative agreement with those of the relativistic 1D code.
However, this issue will eventually be fixed once the relativistic $N$-body code is fully implemented, as this code will evolve the
particle ensemble according to the full metric in three dimensions.

To obtain our estimates for the power spectra of $\Phi-\Psi$, $B_i$ and $h_{ij}$ we ran a suite of $N$-body simulations with the public code \textsc{Gadget-2}~\cite{Springel:2000yr,Springel:2005mi}.
Each of the $12$ simulations contains $(512)^3$ particles, but we ran with various box sizes and resolutions in order to cover a large range of scales. For each realization we make a conservative cut in the
power spectrum at $25\%$ of the Nyquist frequency in order to suppress lattice effects. Modes with wavelengths of more that $12\%$ of the respective box size are also discarded because they may
be sensitive to the finite size of the system. 
We show in \fref{fig:3Dspectra} the power spectra of the metric perturbations, with the error bars representing the realization scatter of the power spectrum estimators, and in \fref{fig:slice} the
real-space representation of the perturbations in a cross section of a simulation snapshot. The box size in that simulation was $320$ Mpc$/h$
and the snapshot was taken at $z=1$.

In \fref{fig:3Dspectra} we can see that the scalar anisotropic stress and the gravitational waves induced by the non-relativistic matter perturbations are small,
about four or five orders of magnitude below the amplitude of the scalar metric perturbations themselves. The spin-1 perturbation $B_i$
is much larger with almost $1\%$ of the amplitude of the scalar potentials (a factor $\sim 10^{-5}$ -- $10^{-4}$ in power). This happens because $B_i$ is sourced by vorticity, which is
only suppressed by one power of $v/c$ over the density, whereas anisotropic stress suffers quadratic suppression. The disparity is therefore owed to the non-relativistic nature
of dark matter, which is the only source of stress-energy perturbations considered in this simple setup. Indeed, when relativistic sources are considered, similar amplitudes of
spin-0, spin-1 and spin-2 perturbations are possible, e.g.\ in the case of topological defects \cite{Durrer:2001cg}.

We also compare our results to analytic calculations from relativistic second order perturbation theory, see, e.g.~\cite{Acquaviva:2002ud,Ananda:2006af,Lu:2008ju,Baumann:2007zm}.
For the vector power spectrum we take the formulas from \cite{Lu:2008ju} and evaluate them with the parametrization of the linear transfer function \cite{Eisenstein:1997ik} used
to generate the initial conditions for our Newtonian $N$-body simulations. As can be seen in \fref{fig:3Dspectra}, the perturbative calculation agrees perfectly with the numerical
results on linear scales, as should be expected. The differences at nonlinear scales seem to be consistent with the results of \cite{Bruni:2013mua}.

A similar calculation can be done for the power spectrum of spin-2 perturbations.
Again we see that it provides an exquisite fit to the numerical results on all scales where perturbations are still linear.
Some details on the perturbative analysis are given in \ref{s:pertspectra}. It is interesting to note that for low values of the inflationary
tensor-to-scalar ratio, $r \lesssim 0.1$, one expects that the tensor perturbations induced by structure formation can be the dominant spin-2
contribution at large sub-horizon scales \cite{Baumann:2007zm}.

If one tries to carry out the perturbative calculation for the scalar term $\Phi-\Psi$, one finds that the corresponding convolution integral is divergent in the infrared for any primordial scalar spectral
index $n_s \leq 1$. This is not very surprising: the perturbative power spectrum of $\Phi^2$ for instance, which is of the same post-Newtonian order, is infrared divergent in a similar way.
At this point one can either take the view that perturbation theory does not give a prediction for this quantity, or one has to impose a regulator to render the convolution integral finite.
The simplest solution is to just neglect the contribution to the convolution integral from $k \lesssim {\cal H}$ since fluctuations on scales larger than the Hubble scale cannot be distinguished
from a contribution to the background. An analytic result for this choice of cutoff is plotted in \fref{fig:3Dspectra}. We notice that the numerical power estimate lies significantly
below the curve on small scales. The reason is quite mundane: the spectra at short wavelengths are obtained from simulations which cover a range of scales at corresponding high wave numbers.
These simulations have relatively small box sizes, and it is the box size with effectively acts as the regulating cutoff. If one would take this into account, each individual simulation
would be fully consistent with the analytic prediction on all scales where it should be valid.

\section{Conclusions and outlook}\label{sec:outlook}

We have described a method for $N$-body simulations in cosmology which smoothly connects relativistic linear perturbation theory to Newtonian $N$-body simulations
and includes additional corrections which can become relevant on intermediate scales or in the presence of evolving relativistic sources. The method is valid as long as
metric perturbations are small, i.e.\ far away from black holes and neutron stars. We have shown that for the standard $\Lambda$CDM
Universe the corrections are expected to remain small. The most relevant term comes from frame-dragging which can amount to changes in the metric perturbations of up to 1\%.
For now, these results have been obtained, somewhat inconsistently, from Newtonian $N$-body simulations. 
We have also verified that the results are in qualitative agreement with our previous relativistic 1D simulations.
In the future they will be tested consistently with the relativistic $N$-body method, once a
3D code is fully implemented.

Although a complete implementation of the relativistic $N$-body scheme is
not yet accomplished at the time of this writing, 
code development is already underway. A first step will be to run an ensemble of 3D simulations of structure
formation for a baseline $\Lambda$CDM cosmology. On the one hand, this will be useful for code validation, because we can compare
numerical results to post-Newtonian estimates or perturbative calculations. On the other hand, such simulations will contain
the most relevant relativistic corrections which
may become important for the next generation of large surveys like Euclid, SKA and LSST.

The most interesting prospects, however, may come from applications beyond $\Lambda$CDM, in particular those where
relativistic contributions to the stress-energy tensor are relevant. We think that the ability to study such scenarios in
detail will justify the additional expense in terms of computational resources {required} by relativistic $N$-body methods.
This extra effort is unavoidable if one is interested in the full dynamical content of general relativity, for instance if one wants to use cosmology to test
our theory of gravity. That the expense can be
mastered and that it is possible to perform self-consistent relativistic simulations of cosmological structure formation
is one of the things we hope to have demonstrated here.

\section*{Acknowledgments}

We would like to thank Marco Bruni for helpful correspondence concerning the post-Newtonian approach.
The $N$-body simulations from which we extracted the post-Newtonian estimates were run on the Andromeda-cluster at Universit\'e de Gen\`eve using the public
code \textsc{Gadget-2}, on which we received useful advice from Vincent Desjacques. We acknowledge financial support from the Swiss National Science Foundation.

\appendix

\section{Green's function method for the gravitational wave power spectrum}\label{s:gws}
If we only want to know the power spectrum or the unequal time correlator of the gravitational waves,
we can use the Green's function method, which only requires the unequal time correlator of the source as an input. We define
\begin{equation}
\label{e:hsource}
 \langle T_{(+2)}(\mathbf{k},\tau)T^*_{(+2)}(\mathbf{k'},\tau')\rangle =(2\pi)^3\delta^3(\mathbf{k}-\mathbf{k'}) 
S(\tau,\tau',k) \, ,
\end{equation}
and 
\begin{equation}
\label{e:hpower}
 2\langle h_{(+2)}(\mathbf{k},\tau_1)h^*_{(+2)}(\mathbf{k'},\tau_2)\rangle =(2\pi)^3\delta^3(\mathbf{k}-\mathbf{k'}) 
P_h(\tau_1,\tau_2,k) \, .
\end{equation}
Here $T_{(+2)}=e_{+2}^{ij}T_{ij}$ and  $h_{(+2)}=e_{+2}^{ij}h_{ij}$ are the helicity-$2$ components of the stress tensor and the metric perturbation, respectively. The function 
$P_h(\tau_1,\tau_2,k)$ is the unequal time correlator in Fourier space while $P_h(\tau,\tau,k) \equiv P_h(\tau,k)$ is the power spectrum.
The factor 2 accounts for the  contributions of the two helicities which we assume to have identical spectra and to be uncorrelated. The fact that the evolution equation for $h_{ij}$ is linear implies
\begin{equation}
\label{e:hsol}
P_h(\tau_1,\tau_2,k) =2\left(\frac{16\pi G}{k^2}\right)^2\int_{\tau_{\rm in}}^\tau d\tau'd\tau''{\cal G}(\tau_1,\tau',k){\cal G}^*(\tau_2,\tau'',k)S(\tau',\tau'',k)\,,
\end{equation}
where ${\cal G}(\tau,\tau',k)$ is the Green's function for \eref{e:hmode}. In terms of the homogeneous solutions, $h_1(k\tau),~h_2(k\tau)$, it can be expressed as
$$
{\cal G}(\tau,\tau',k)= k\frac{h_1(k\tau)h_2(k\tau')-h_1(k\tau')h_2(k\tau)}{h'_1(k\tau')h_2(k\tau')-h_1(k\tau')h'_2(k\tau')} =k\frac{h_1(k\tau)h_2(k\tau')-h_1(k\tau')h_2(k\tau)}{w(k\tau')}\,.
$$
 In the matter dominated era the homogeneous solutions are given by spherical Bessel functions, $h_1(x)=j_1(x)/x$ and $h_2(x)=y_1(x)/x$ with Wronskian $w(x)=x^{-4}$,
 see e.g.~\cite{2008cmbg.book}.
 In the era where dark energy is relevant, the homogeneous solution is no longer simply a function of $x=k\tau$ and it has to be determined numerically.
 
\section{Perturbative formulas for the vector and tensor power spectra induced by dark matter at second order}\label{s:pertspectra}

In this appendix we summarize the perturbative calculation of the vector and tensor power spectra. Since they appear only at second order, they are essentially obtained as squares of
first order terms. At first order, we can set $\Phi = \Psi$, and we can use the linearized Einstein's equations to connect the first order density and velocity perturbation
to the scalar potential,
\begin{eqnarray}
 \delta_{ij} v^j &=& - \frac{2}{3 \mathcal{H}^2 \Omega_\mathrm{m}} \left(\Psi_{,i}' + \mathcal{H} \Psi_{,i}\right) \, ,\\
 \frac{\delta\rho}{\rho} &=& \frac{2}{3 \mathcal{H}^2 \Omega_\mathrm{m}} \left(\Delta \Psi - 3 \mathcal{H} \Psi' - 3 \mathcal{H}^2 \Psi\right) \, .
\end{eqnarray}
Substituting these expressions into \eref{e:G0i} and going to Fourier space, we obtain at lowest non-trivial order
\begin{equation}
 k^2 B_i(\mathbf{k}) = \frac{8 \rmi}{3 \mathcal{H}^2 \Omega_\mathrm{m}} \left(2 \pi\right)^{-\frac{3}{2}} \int\!\rmd^3 \mathbf{q} P_{ij} q^2 \left(k^j - q^j\right) \Psi(\mathbf{q}) \left[\Psi'(\mathbf{k}-\mathbf{q}) + \mathcal{H} \Psi(\mathbf{k}-\mathbf{q})\right] \, ,
\end{equation}
where $P_{ij} \doteq \delta_{ij} - k_i k_j / k^2$ is the transverse projector. The calculation further simplifies if we specify $\Psi(\mathbf{k})$ in terms of some initial
value, the linear transfer function $T(k)$ and the growth factor $g(z)$ as $\Psi(\mathbf{k}) = g T(k) \lim_{z\rightarrow\infty}\Psi(\mathbf{k})$. The power spectrum of $B_i$
can then be expressed in terms of the primordial power spectrum of $\Psi$,
\begin{eqnarray}
 \fl P_{B}(k) = \int\!\rmd^3 \mathbf{q} \left(2 \delta_{ij} q^i k^j - k^2\right) \left(q^2 - \frac{\left(\delta_{ij} q^i k^j\right)^2}{k^2}\right) q^2 T^2(q) T^2(|\mathbf{k}-\mathbf{q}|) P^\mathrm{in}_\Psi(q) P^\mathrm{in}_\Psi (|\mathbf{k}-\mathbf{q}|) \nonumber\\
 \times \frac{64 k^{-4}}{9 \mathcal{H}^2 \Omega_\mathrm{m}^2} g^2 \left[g - \left(1+z\right) \frac{\rmd g}{\rmd z}\right]^2 \, .
\end{eqnarray}
We used the common assumption that the initial conditions for the first order potential are Gaussian.

A similar calculation can be done for $h_{ij}$. In this case we use \eref{e:Gijsw} which, after neglecting the time derivatives on $h_{ij}$ (quasistatic limit)
and going again to Fourier space becomes
\begin{eqnarray}
 \fl\frac{1}{2} k^2 h_{ij}(\mathbf{k}) = \left(2 \pi\right)^{-\frac{3}{2}} \int\!\rmd^3 \mathbf{q} \Lambda_{ijlm} \Biggl\{ 2 \left(q^l q^m - \frac{1}{3}\delta^{lm} q^2\right) \Psi(\mathbf{q}) \Psi(\mathbf{k}-\mathbf{q})\nonumber\\
  -\frac{4}{3 \mathcal{H}^2 \Omega_\mathrm{m}} \left(q^l k^m - q^l q^m - \frac{1}{3} \delta^{lm} \delta_{rs} q^r k^s + \frac{1}{3} \delta^{lm} q^2\right)\nonumber\\
 \times \left[\Psi'(\mathbf{q}) + \mathcal{H} \Psi(\mathbf{q})\right] \left[\Psi'(\mathbf{k}-\mathbf{q}) + \mathcal{H} \Psi(\mathbf{k}-\mathbf{q})\right]\Biggr\} \, ,
\end{eqnarray}
where $\Lambda_{ijlm} \doteq P_{il} P_{jm} - \frac{1}{2} P_{ij} P_{lm}$ is the projector on the spin-2 component. Applying the same treatment as in the previous case, the power spectrum
of $h_{ij}$ is expressed as
\begin{eqnarray}
 \fl P_h(k) = \int\!\rmd^3 \mathbf{q} \left(q^2 - \frac{\left(\delta_{ij} q^i k^j\right)^2}{k^2}\right)^2 T^2(q) T^2(|\mathbf{k}-\mathbf{q}|) P^\mathrm{in}_\Psi(q) P^\mathrm{in}_\Psi (|\mathbf{k}-\mathbf{q}|) \nonumber\\
 \times \left[4 g^2 + \frac{8}{3 \Omega_\mathrm{m}} \left(g - \left(1+z\right)\frac{\rmd g}{\rmd z}\right)^2\right]^2 k^{-4} \, .
\end{eqnarray}

We notice that for $k\rightarrow 0$ the power in the tensor fluctuations behaves like $k^3 P_h(k) \sim 1/k$. This behaviour
is already visible on large scales in \fref{fig:3Dspectra}. However, the apparent divergence is due to the quasistatic approximation
which eventually breaks down outside the horizon. A calculation based on the Green's function method reveals that the spectral tilt
outside the horizon is given by $k^3 P_h(k) \sim k^3$, see e.g.\ \cite{Baumann:2007zm}. This can be understood from the fact that induced second order fluctuations are causal and hence lead to a white noise spectrum  $P_h$ on super horizon scales.
The scales shown in \fref{fig:3Dspectra} are all inside the horizon where the quasistatic limit gives a reasonable approximation.

\section*{References}

\bibliographystyle{iopart-num}
\bibliography{julian}

\end{document}